\documentclass[12pt]{iopart}
\begin{document}

\title[Anharmonic oscillators]{Quantum anharmonic oscillators: a new approach}
\author{F J G\'omez and J Sesma}
\address{Departamento de F\'{\i}sica Te\'orica, Facultad de
Ciencias, 50009, Zaragoza, Spain}

\begin{abstract}
The determination of the eigenenergies of a quantum anharmonic
oscillator consists merely in finding the zeros of a function of
the energy, namely the Wronskian of two solutions of the
Schr\"odinger equation which are regular respectively at the
origin and at infinity. We show in this paper how to evaluate that
Wronskian exactly, except for numerical rounding errors. The
procedure is illustrated by application to the $gx^2+x^{2N}$ ($N$
a positive integer) oscillator.
\end{abstract}

\pacno{03.65.Ge}

\submitto{\JPA}

\maketitle

\nosections

Quantum anharmonic oscillators have been frequently used in
different branches of Physics to simulate a great variety of
situations and to explain multitude of phenomena. Apart from this,
since the publication of the seminal papers by Bender and Wu
\cite{bend} and by Simon and Dicke \cite{simo} showing the failure
of the Rayleigh-Schr\"odinger perturbation method, they have
served to test plenty of approximate methods of solution of the
Schr\"odinger equation. Papers dealing with the most recently
proposed methods \cite{meth} contain references to older ones,
that we omit for brevity. It seems, however, to have been passed
unnoticed that an exact procedure exists to obtain a quantization
condition that gives the eigenenergies as zeros of an easily
calculable function. The idea is the same one exploited in the
solution of the Schr\"odinger equation for the harmonic oscillator
or the Coulomb potential, although the procedure is slightly more
tricky than in those two simple cases.

The starting point is the Schr\"odinger equation written as a
second order differential equation free of first-derivative terms.
Such equation is satisfied by the wave function, in the case of an
even one-dimensional anharmonic oscillator, or by the reduced wave
function, if one is considering an isotropic D-dimensional
oscillator. A solution $u{\scriptstyle {\rm {reg}}}$ of the
differential equation physically acceptable at the origin can be
immediately obtained as a power series of the variable. Two other
solutions $u^{(1)}$ and $u^{(2)}$, characterized by their behavior
at large distances, can also be considered. To be specific, let
$u^{(1)}$ represent the solution going exponentially to zero as
the variable increases, whereas $u^{(2)}$ corresponds to an
exponentially diverging one. Since $u^{(1)}$ and $u^{(2)}$ are
independent, $u{\scriptstyle {\rm {reg}}}$ can always be written
as a linear combination of them with coefficients, called {\em
connection factors}, which depend on the energy and on the
parameters of the potential. For a generic value of the energy,
both connection factors are different from zero and
$u{\scriptstyle {\rm {reg}}}$ is not a physical solution because
of its behavior at infinity. The eigenenergies are then determined
by requiring the cancellation of the connection factor multiplying
$u^{(2)}$.

The connection problem for a differential equation with two
singular points (let us say, one at the origin and the other at
infinity) was discussed by Naundorf \cite{nau}. He considered the
case of one of the singular points (that at infinity, for
instance) being irregular of integer rank $R>0$ and the other one
being either irregular of integer rank $r>0$ or regular ($r=0$).
Here we are interested in this last case. Naundorf gave a
procedure consisting in obtaining $2R$ independent formal power
series, with an integer index running from $-\infty$ to $+\infty$,
having well defined asymptotic behaviors, and whose coefficients
can serve as a basis in the $2R$-dimensional space of solutions of
the recurrence obeyed by the coefficients of $u_{\scriptstyle {\rm
{reg}}}$. To obtain such basis, Naundorf replaces, in the known
expressions of $u^{(1)}$ and $u^{(2)}$ \cite{olv}, the exponential
term determining their respective asymptotic behaviors by $R$
independent formal expansions of the type of the Heaviside's
exponential series. Multiplication of those formal expansions by
the Taylor series of the rest of the exponential terms and the
descending power series in $u^{(1)}$ and $u^{(2)}$ produces $2R$
formal expansions whose coefficients obey the above mentioned
recurrence, i. e., the required basis. Comparison of $2R$
consecutive coefficients of the power series expression of
$u_{\scriptstyle {\rm {reg}}}$ with the analogous coefficients of
the elements of the basis leads to a system of $2R$ linear
equations whose solution allows one to obtain the connection
factors. That procedure has been applied to the solution of
several physical problems, like the hydrogen atom with fine
structure \cite{nau2}, the quarkonium \cite{gom1}, the spherical
Stark effect in the hydrogen atom \cite{gom2} or the quartic and
sextic anharmonic oscillators \cite{gom3}.

The method here suggested is related to the Naundorf's one insofar
as it also rests on the vanishing of one of the connection factors
and makes use of Heaviside's exponential series to obtain formal
expansions, but differs from the Naundorf's method in the
procedure of computation: instead of following the steps detailed
in the preceding paragraph, we benefit from the fact that the
connection factor multiplying $u^{(2)}$ is given by the quotient
of Wronskians $\mathcal{W}[u_{\scriptstyle {\rm
{reg}}},u^{(1)}]/\mathcal{W}[u^{(2)},u^{(1)}]$ and, since the
denominator does not vanish, the quantization results from the
fulfillment of the condition
\begin{equation}
\mathcal{W}[u_{\scriptstyle {\rm {reg}}},u^{(1)}]=0. \label{uno}
\end{equation}
To implement this condition, we need suitable expressions of
$u_{\scriptstyle {\rm {reg}}}$ and $u^{(1)}$. The series expansion
mentioned above is adequate to represent the first of these
solutions. For the second one a closed expression does not exist,
in general, but a formal (asymptotic) expansion can be easily
obtained by substitution in the differential equation. Then it is
trivial to write a formal expression of the Wronskian and to
require its cancellation.

To illustrate the method, let us apply it to the determination of
the eigenenergies of the one-dimensional anharmonic oscillator
represented by the potential
\begin{equation}
V(x)=g\,x^2+x^{2N},   \qquad N \; \mbox{a positive integer}.
\label{dos}
\end{equation}
This problem has been tackled by several authors
\cite{guar,nana,amor} by using different approximations. We
discard the trivial case $N=1$. The cases $N=2$ (usually referred
to as quartic oscillator) and $N=3$ (sextic) can be easily solved
following the steps we are going to detail, but the resulting
equations do not fit in the general form given below. Therefore,
we assume $N\geq 4$. The Schr\"odinger equation (in adequate units
for the variable $x$ and the energy $E$)
\begin{equation}
\left( -\frac{d^2}{dx^2}+g\,x^2+x^{2N}\right) u(x)=E\,u(x),
\label{tres}
\end{equation}
admits solutions, regular at the origin, of the form
\begin{equation}
u_{\scriptstyle {\rm {reg}}}(x)=\sum_{n=0}^\infty
a_n\,x^{n+\nu},\qquad a_0\neq 0, \label{cuatro}
\end{equation}
with $\nu =0$ (even states) or $1$ (odd states). Alternatively,
two independent solutions, $u^{(1)}$ and $u^{(2)}$, with
asymptotic expansions (for $x\to +\infty$)
\begin{equation}
u^{(j)}(x)\sim\exp\left[\frac{\alpha^{(j)}}{N\! +\!
1}\,x^{N+1}\right] x^{\mu^{(j)}}\sum_{m=0}^\infty
h_m^{(j)}\,x^{-m}, \quad h_0^{(j)}\neq 0, \quad j=1, 2,
\label{cinco}
\end{equation}
can also be considered. Substitution of this formal expansion in
(\ref{tres}) gives for the exponents
\begin{equation}
\begin{array}{ll}
\alpha^{(1)}=-1,  \qquad & \mu^{(1)} =\mu \equiv -N/2, \\
\alpha^{(2)}=+1, \qquad &  \mu^{(2)} =\mu \equiv -N/2,
\end{array}  \label{seis}
\end{equation}
and for the coefficients
\begin{equation}
2\alpha^{(j)}m\,h_m^{(j)}=(m\! -\! N/2)(m\! -\! N/2\! -\! 1)\,
h_{m-N-1}^{(j)}+E\,h_{m-N+1}^{(j)} -g\,h_{m-N+3}^{(j)}.
\label{siete}
\end{equation}
Instead of computing directly the left hand side of (\ref{uno}),
let us introduce two auxiliary functions
\begin{eqnarray}
v_{\scriptstyle {\rm {reg}}}(x) & = & \exp\left(
x^{N+1}/(N+1)\right)\,u_{\scriptstyle {\rm {reg}}}(x),
\label{ocho} \\ v^{(1)}(x) & = & \exp\left(
x^{N+1}/(N+1)\right)\,u^{(1)}(x), \label{nueve}
\end{eqnarray}
which obey the differential equation
\begin{equation}
\frac{d^2v}{dx^2}-2\,x^N\,\frac{dv}{dx}+\left(
E-g\,x^2-Nx^{N-1}\right) v = 0,  \label{diez}
\end{equation}
and whose Wronskian satisfies
\begin{equation}
\mathcal{W}[v_{\scriptstyle {\rm {reg}}},v^{(1)}]=\exp\left(
2x^{N+1}/(N+1)\right)\,\mathcal{W}[u_{\scriptstyle {\rm
{reg}}},u^{(1)}].  \label{duno}
\end{equation}
Now, by using the series expansion
\begin{equation}
v_{\scriptstyle {\rm {reg}}}(x)=\sum_{n=0}^\infty b_n\,x^{n+\nu},
\qquad b_0\neq 0, \label{ddos}
\end{equation}
with coefficients given by the recurrence
\begin{equation}
(n+\nu )(n+\nu
-1)\,b_n=-E\,b_{n-2}+g\,b_{n-4}+2(n-N/2-1+\nu)\,b_{n-N-1},
\label{dtres}
\end{equation}
and the asymptotic expansion
\begin{equation}
v^{(1)}(x)\sim\sum_{m=0}^{\infty}
h_m^{(1)}\,x^{-m+\mu},\label{dcuatro}
\end{equation}
one obtains for the left hand side of (\ref{duno}) a formal
expansion
\begin{equation}
\mathcal{W}[v_{\scriptstyle {\rm
{reg}}},v^{(1)}]\sim\sum_{k=-\infty}^{\infty}\gamma_k\,x^{k-1+\nu+\mu},
\label{dcinco}
\end{equation}
with coefficients
\begin{equation}
\gamma_k = \sum_{m=0}^{\infty}(-2m-k-\nu+\mu)\,b_{m+k}\,h_m.
\label{dseis}
\end{equation}
A similar expansion can be obtained for the right hand side of
(\ref{duno}) by recalling the Heaviside's exponential series
\begin{equation}
\exp(t)\sim\sum_{n=-\infty}^{\infty}\frac{t^{n+\delta}}{\Gamma(n+1+\delta)},
\label{dsiete}
\end{equation}
introduced by Heaviside in the second volume of his {\em
Electromagnetic theory} (London, 1899) and probed by Barnes
\cite{bar} to be an asymptotic expansion for arbitrary $\delta$
and $|\arg (t)|<\pi$. Let us construct $N+1$ expansions
\begin{equation}
\exp\left( 2x^{N+1}/(N+1)\right)\sim \mathcal{E}_L
\equiv\sum_{n=-\infty}^{\infty}
\frac{(2x^{N+1}/(N+1))^{n+\delta_L}}{\Gamma(n+1+\delta_L)}
  \label{docho}
\end{equation}
of the type (\ref{dsiete}) with appropriate choices for $\delta$,
\begin{equation}
\delta_L = (\nu +\mu +L)/(N+1),\qquad L=0, 1, \ldots , N.
\label{dnueve}
\end{equation}
It is evident that, for any set of constants $\beta_L$ ($L=0, 1,
\ldots, N$) restricted by the condition
\begin{equation}
\mathcal{W}[u_{\scriptstyle {\rm
{reg}}},u^{(1)}]=\sum_{L=0}^N\beta_L, \label{veinte}
\end{equation}
one has
\begin{equation}
\exp\left( 2x^{N+1}/(N+1)\right)\mathcal{W}[u_{\scriptstyle {\rm
{reg}}},u^{(1)}]\sim \sum_{L=0}^N \beta_L\,\mathcal{E}_L.
\label{vuno}
\end{equation}
If, according to Eq. (\ref{duno}), this formal expansion has to
coincide with that in (\ref{dcinco}), the constants $\beta_L$ must
be
\begin{equation}
\beta_L= \frac{\Gamma
(n+1+\delta_L)}{(2/(N+1))^{n+\delta_L}}\,\gamma_{k_L}, \qquad
k_L=n(N+1)+1+L,  \label{vdos}
\end{equation}
where the integer $n$ can be chosen at will. Substitution of those
values in (\ref{veinte}) allows one to write the quantization
condition (\ref{uno}) in the final form
\begin{equation}
\sum_{L=0}^N\Gamma (n+1+\delta_L)\,((N+1)/2)^{L/(N+1)}\,
\gamma_{k_L}= 0. \label{vtres}
\end{equation}

We have used the last expression of the quantization condition to
find the lowest eigen\-energies of the anharmonic oscillator
(\ref{dos}) for different values of the coupling parameter $g$ and
four different  choices of $N$. In the computation we have used a
FORTRAN program with double precision. The results are shown in
Tables 1 to 4.

\begin{table}
\caption{The four lowest eigenenergies of the oscillator
(\ref{dos}), for $N=4$ and several values of $g$. The energies
$E_0$ and $E_2$ correspond to even states ($\nu=0$ in
(\ref{cuatro})); $E_1$ and $E_3$ to odd ones ($\nu=1$).}
\begin{indented}
\item[]
\begin{tabular}{rrrrr}
\br
 $g$ & $E_0$\hspace{0.5 cm} & $E_1$\hspace{0.5 cm}  &
$E_2$\hspace{0.5 cm} & $E_3$\hspace{0.5 cm} \\
\mr
 --\,20& \quad --\, 15.62781790 & $\;$ --\, 15.60342843 &
$\;$ --\, 1.99759674 & $\;$ 0.04913769 \\
--\,10 & \quad --\,3.89894214 & $\;$ --\, 3.32541335 & $\;$
3.26415045 & $\;$ 8.82212629 \\
--\,1 & \quad 0.93527862 & $\;$ 4.11346827 & $\;$ 9.49008984 &
$\;$ 16.49163253 \\
--\,0.1 & \quad 1.19798114 & $\;$ 4.69299658 & $\;$
10.16968229 & $\;$ 17.25807961 \\
0 & \quad 1.22582011 & $\;$ 4.75587441 & $\;$
10.24494698 & $\;$ 17.34308797 \\
0.1 & \quad 1.25340643 & $\;$ 4.81845727 & $\;$
10.32015025 & $\;$ 17.42806187 \\
1 & \quad 1.49101990 & $\;$ 5.36877806 & $\;$
10.99373734 & $\;$ 18.19110002 \\
10 & \quad 3.21296474 & $\;$ 9.86889192 & $\;$
17.20002166 & $\;$ 25.52311499 \\
20 & \quad 4.48741520 & $\;$ 13.54543209 & $\;$
22.89430780 & $\;$ 32.78247104 \\
\br
\end{tabular}
\end{indented}
\end{table}

\begin{table}
\caption{The four lowest eigenenergies of the Hamiltonian
(\ref{dos}), for $N=5$ and several values of $g$.}
\begin{indented}
\item[]
\begin{tabular}{rrrrr}
\br
 $g$ & $E_0$\hspace{0.5 cm} & $E_1$\hspace{0.5 cm}  &
$E_2$\hspace{0.5 cm} & $E_3$\hspace{0.5 cm} \\
\mr
 --\,20& \quad --\, 11.56630147 & $\;$ --\, 11.45854677 &
$\;$  0.56494700 & $\;$ 4.90729085 \\
--\,10 & \quad --\, 2.83782675 & $\;$ --\, 1.83075483 & $\;$
4.90946147 & $\;$ 11.94279256 \\
--\,1 & \quad 1.03205834 & $\;$ 4.51533389 & $\;$ 10.48697985 &
$\;$ 18.45464482 \\
--\,0.1 & \quad 1.27308185 & $\;$ 5.04058836 & $\;$ 11.08762465
 & $\;$ 19.11537634 \\
0 & \quad 1.29884370 & $\;$ 5.09787653 & $\;$
11.15431820 & $\;$ 19.18880956 \\
0.1 & \quad 1.32441224 & $\;$ 5.15495387 & $\;$
11.22099452 & $\;$ 19.26224408 \\
1 & \quad 1.54626351 & $\;$ 5.65933772 & $\;$
11.81996788 & $\;$ 19.92310357 \\
10 & \quad 3.21711708 & $\;$ 9.93229322 & $\;$
17.51589563 & $\;$ 26.43450876 \\
20 & \quad 4.48623513 & $\;$ 13.55329264 & $\;$
22.99231828 & $\;$ 33.19354764 \\
\br
\end{tabular}
\end{indented}
\end{table}

\begin{table}
\caption{The four lowest eigenenergies of the Hamiltonian
(\ref{dos}), for $N=6$ and several values of $g$.}
\begin{indented}
\item[]
\begin{tabular}{rrrrr}
\br
 $g$ & $E_0$\hspace{0.5 cm} & $E_1$\hspace{0.5 cm}  &
$E_2$\hspace{0.5 cm} & $E_3$\hspace{0.5 cm} \\
\mr
 --\,20& \quad --\, 9.36607177 & $\;$ --\, 9.13010587 &
$\;$ 2.01035459 & $\;$ 7.97554684 \\
--\,10 & \quad --\, 2.24187409 & $\;$ --\, 0.87004433 & $\;$
6.12159677 & $\;$ 14.16512836 \\
--\,1 & \quad 1.11369983 & $\;$ 4.84470202 & $\;$ 11.28130698 &
$\;$ 19.99987959 \\
--\,0.1 & \quad 1.33949907 & $\;$ 5.33347217 & $\;$
11.83181276 & $\;$ 20.59539382 \\
0 & \quad 1.36377971 & $\;$ 5.38694202 & $\;$
11.89300908 & $\;$ 20.66163760 \\
0.1 & \quad 1.38786579 & $\;$ 5.44024556 & $\;$
11.95420520 & $\;$ 20.72789495 \\
1 & \quad 1.59799050 & $\;$ 5.91264617 & $\;$
12.50470842 & $\;$ 21.32474109 \\
10 & \quad 3.22441873 & $\;$ 10.00630419 & $\;$
17.83164730 & $\;$ 27.27876498 \\
20 & \quad 4.48680192 & $\;$ 13.57082013 & $\;$ 23.11371663
& $\;$ 33.63281210 \\
\br
\end{tabular}
\end{indented}
\end{table}

\begin{table}
\caption{The four lowest eigenenergies of the oscillator
(\ref{dos}), for $N=7$ and several values of $g$.}
\begin{indented}
\item[]
\begin{tabular}{rrrrr}
\br
 $g$ & $E_0$\hspace{0.5 cm} & $E_1$\hspace{0.5 cm}  &
$E_2$\hspace{0.5 cm} & $E_3$\hspace{0.5 cm} \\
\mr
 --\,20& \quad --\, 7.97489149 & $\;$ --\, 7.59026706 &
$\;$ 3.05916112 & $\;$ 10.19269195 \\
--\,10 & \quad --\, 1.8474624 & $\;$ --\, 0.17159144 & $\;$
7.07320094 & $\;$ 15.87259291 \\
--\,1 & \quad 1.18393765 & $\;$ 5.12329191 & $\;$ 11.93911991 &
$\;$ 21.26204013 \\
--\,0.1 & \quad 1.39832030 & $\;$ 5.58552094 & $\;$
12.45475050 & $\;$ 21.81341553 \\
0 & \quad 1.42143888 & $\;$ 5.63618503 & $\;$
12.51210199 & $\;$ 21.87477520 \\
0.1 & \quad 1.44442247 & $\;$ 5.68671175 & $\;$
12.56946066 & $\;$ 21.93615283 \\
1 & \quad 1.64542730 & $\;$ 6.13534277 & $\;$
13.08581400 & $\;$ 22.48930458 \\
10 & \quad 3.23335919 & $\;$ 10.08415888 & $\;$
18.13465608 & $\;$ 28.04433038 \\
20 & \quad 4.48835326 & $\;$ 13.59428939 & $\;$ 23.24781210
& $\;$ 34.07417453 \\
\br
\end{tabular}
\end{indented}
\end{table}

The procedure presented above assumes the capability to compute
the $N+1$ coefficients $\gamma_{k_L}$ ($L=0, 1, \ldots , N$) by
summation of the series in (\ref{dseis}). We have not yet proved
rigorously the convergence of such series, albeit extensive
numerical explorations guarantee its convergence for sufficiently
large $k$, i. e., for $n$, in Eq. (\ref{vdos}), above a certain
threshold which depends on the values of the coupling parameter
$g$ and on the energy. Moreover, those explorations show that, the
larger $n$ is taken, the faster becomes the convergence.
Investigations tending to elucidate that question are currently in
progress.

Besides the eigenenergies, our method determines also, in
principle, the eigenfunctions. In the example considered, they are
given by Eqs. (\ref{ocho}) and (\ref{ddos}). Nevertheless,
although the series in (\ref{ddos}) converges for all finite $x$,
it cannot be used safely for large values of $x$, unless a
considerable number of digits are maintained in the successive
arithmetical operations. Certainly, the asymptotic expansion
(\ref{cinco}) can be used for sufficiently large values of $x$
(above about 5 units). But it is not clear the advantage of this
procedure over the conventional numerical integration of the
Schr\"odinger equation, especially if one needs the normalized
wave function for a large number of points.

To facilitate understanding the method, we have chosen above a
very simple example: a one-dimensional anharmonic oscillator with
only two terms in the potential. The procedure is equally
applicable to isotropic D-dimensional oscillators with any number
of integer powers of the radial variable in the potential and for
any value of the D-dimensional ``angular momentum". It is also
applicable, of course, to easier problems. Let us consider, for
instance, three anharmonic oscillators algebraically solvable,
namely, the P\"oschl-Teller, the modified P\"oschl-Teller and the
Morse potentials. Their exact solution can be found in Ref.
\cite{flu}, whose notation we adopt. In what follows, we
concentrate on obtaining, by our procedure, the eigenenergies of
the bound states. But, since the reflection and transmission
coefficients are trivially related to the connection factors, our
method is also useful for calculating phase shifts.

In the case of the P\"oschl-Teller potential
\begin{equation}
V(x)=\frac{1}{2}V_0\left(\frac{\kappa (\kappa -1)}{\sin^2(\alpha
x)} + \frac{\lambda (\lambda -1)}{\cos^2(\alpha x)}\right), \qquad
V_0=\frac{\hbar^2\alpha^2}{m}, \quad \kappa, \lambda >1,
\label{vcuatro}
\end{equation}
defined in the interval $x\in [0,\pi/2]$, the Schr\"odinger
equation can be written in the form
\begin{equation}
y(y-1)u^{\prime\prime}+\left(\frac{1}{2}-y\right)u^{\prime} +
\frac{1}{4}\left(\frac{k^2}{\alpha^2}-\frac{\kappa (\kappa -1)}{y}
- \frac{\lambda (\lambda -1)}{1-y}\right)u = 0, \label{vcinco}
\end{equation}
in terms of a new variable
\begin{equation}
y=\sin^2(\alpha x)  \label{vseis}
\end{equation}
and using, instead of the energy $E$, the parameter
\begin{equation}
k^2=\frac{2mE}{\hbar^2}.  \label{vsiete}
\end{equation}
Equation (\ref{vcinco}) can be written as a hypergeometric one by
means of the change of function done in Ref. \cite{flu}. Then, it
is immediate to write the connection factors and to obtain the
quantization condition. Nevertheless, let us ignore this fact and
try to apply our method directly to Eq. (\ref{vcinco}), to be
solved between the two regular singular points $y=0$ and $y=1$.
The solution regular at $y=0$ can be written as a power series
\begin{equation}
u_{\scriptstyle {\rm {reg}}}(y)=\sum_{n=0}^{\infty}a_n\,
y^{n+\kappa/2}, \qquad a_0\neq 0, \label{vocho}
\end{equation}
with coefficients given by the recurrence
\begin{eqnarray}
\lefteqn{\hspace{-2cm}
n\left(n\!-\!\frac{1}{2}\!+\!\kappa\right)a_n=\left(\left(
n\!-\!1\!+\!\frac{\kappa}{2}\right)\left(2n\!-\!\frac{5}{2}\!
+\!\kappa\right) - \frac{1}{4}\left(\frac{k^2}{\alpha^2}
+\kappa(\kappa\!-\!1)-
\lambda(\lambda\!-\!1)\right)\right)a_{n-1}}  \nonumber \\
& & \hspace{6cm}\ -
\left(\left(n\!-\!2\!+\!\frac{\kappa}{2}\right)^2-
\frac{k^2}{4\alpha^2}\right)a_{n-2}.  \label{vnueve}
\end{eqnarray}
The solution regular at $y=1$ can be immediately written if one
realizes that the differential equation (\ref{vcinco}) is
invariant under the interchange
\[
\left\{\begin{array}{cc} y \\ \kappa\end{array}\right\}
\longleftrightarrow \left\{\begin{array}{cc}1-y \\ \lambda
\end{array}\right\}
\]
and, therefore,
\begin{equation}
u^{(1)}(y)=\sum_{n=0}^{\infty}b_n\, (1-y)^{n+\lambda/2}, \qquad
b_0\neq 0, \label{treinta}
\end{equation}
with coefficients given now by
\begin{eqnarray}
\lefteqn{\hspace{-2cm}
n\left(n\!-\!\frac{1}{2}\!+\!\lambda\right)b_n=\left(\left(
n\!-\!1\!+\!\frac{\lambda}{2}\right)\left(2n\!-\!\frac{5}{2}\!
+\!\lambda\right) - \frac{1}{4}\left(\frac{k^2}{\alpha^2}
+\lambda(\lambda\!-\!1)-
\kappa(\kappa\!-\!1)\right)\right)b_{n-1}}  \nonumber \\
& & \hspace{6cm}\ -
\left(\left(n\!-\!2\!+\!\frac{\lambda}{2}\right)^2-
\frac{k^2}{4\alpha^2}\right)b_{n-2}.  \label{tuno}
\end{eqnarray}
Both series in Eqs. (\ref{vocho}) and (\ref{treinta}) are
convergent for $y\in (0,1)$. We can, therefore, write a convergent
(not merely formal, as in the problem considered before) expansion
of the Wronskian. At the point $y=1/2$, for instance, one has
\begin{eqnarray}
\lefteqn{\hspace{-1cm}\mathcal{W}[u_{\scriptstyle {\rm
{reg}}},u^{(1)}](y=1/2)=-\frac{1}{2^{(\kappa+\lambda)/2}}
\Bigg(\left(\sum_{n=0}^{\infty}\frac{a_n}{2^n}\right)
\left(\sum_{m=0}^{\infty}\frac{(m+\lambda/2)b_m}{2^{m-1}}\right)}
\nonumber \\
& & \hspace{5cm}\ +
\left(\sum_{n=0}^{\infty}\frac{(n+\kappa/2)a_n}{2^{n-1}}\right)
\left(\sum_{m=0}^{\infty}\frac{b_m}{2^m}\right)\Bigg).
\label{tdos}
\end{eqnarray}
Giving numerical values to $\kappa$ and $\lambda$, one can check
that the Wronskian vanishes whenever
\begin{equation}
\frac{k^2}{\alpha^2}=(\kappa+\lambda+2n)^2, \qquad n=0, 1, 2,
\ldots ,  \label{ttres}
\end{equation}
as it should be.

The modified P\"oschl-Teller potential, defined for $x\in
(-\infty, +\infty)$, reads
\begin{equation}
V(x)=-\frac{\hbar^2}{2m}\,\alpha^2\,\frac{\lambda (\lambda
-1)}{\cosh^2(\alpha x)}, \qquad \lambda >1. \label{tcuatro}
\end{equation}
Instead of the (negative) energy $E$, we use the parameter
\begin{equation}
\kappa^2=\frac{2m(-E)}{\hbar^2}.  \label{tcinco}
\end{equation}
Once again, the Schr\"odinger equation can be written as a
hypergeometric one with adequate changes of variable and function.
The change of variable used in Ref \cite{flu} maps the interval
$(-\infty,+\infty)$ for the variable $x$ onto $[1,+\infty)$ for
the new variable. Here we prefer, however, to make a different
change of variable, namely
\begin{equation}
y=\frac{1}{\cosh^2(\alpha x)},  \label{tseis}
\end{equation}
in order to get the mentioned interval, where the differential
equation has to be solved, mapped onto $[0,1]$. The Schr\"odinger
equation turns then into
\begin{equation}
y^2(1-y)u^{\prime\prime}+y\left(1-\frac{3}{2}y\right)u^{\prime} +
\frac{1}{4}\left(-\frac{\kappa^2}{\alpha^2}+ \lambda (\lambda
-1)y\right)u = 0. \label{tsiete}
\end{equation}
The regular solution at the regular singular point $y=0$ can be
written as a series
\begin{equation}
u_{\scriptstyle {\rm {reg}}}(y)=\sum_{n=0}^{\infty}a_n\,
y^{n+\kappa/(2\alpha)}, \qquad a_0\neq 0, \label{tocho}
\end{equation}
with coefficients obtainable by means of
\begin{equation}
n\left(n\!+\!\frac{\kappa}{\alpha}\right)a_n=\left(\left(
n\!-\!1\!+\!\frac{\kappa}{2\alpha}\right)\left(n\!-\!\frac{1}{2}\!
+\!\frac{\kappa}{2\alpha}\right) -
\frac{\lambda(\lambda\!-\!1)}{4}\right)a_{n-1}. \label{tnueve}
\end{equation}
Now we need to write the well behaved solution at $y=1$. Two
independent expansions in power series of $1-y$ of the form,
\begin{equation}
u(y)=\sum_{n=0}^{\infty}b_n\, (1-y)^{n+\mu}, \qquad b_0\neq 0,
\quad \label{cuarenta}
\end{equation}
with
\[
\mu = 0  \qquad {\rm {or}} \qquad \mu=1/2
\]
and coefficients obeying the recurrence
\begin{eqnarray}
\lefteqn{\hspace{-2cm}
\left(n\!+\!\mu\right)\left(n\!+\!\mu\!-\!\frac{1}{2}\right)b_n=\left(2\left(
n\!-\!1\!+\!\mu\right)^2 +\frac{\kappa^2}{4\alpha^2}
-\frac{\lambda(\lambda\!-\!1)}{4}\right)b_{n-1}}  \nonumber \\
& & \hspace{3cm}\ -
\left(\left(n\!-\!2\!+\mu\right)\left(n\!-\!\frac{3}{2}\!+\mu\right)-
\frac{\lambda(\lambda\!-\!1)}{4}\right)b_{n-2},  \label{cuno}
\end{eqnarray}
are physically acceptable. Those solutions with $\mu=0$ and
$\mu=1/2$ correspond, respectively, to even and odd wavefunctions
in the variable $x$. Choosing the point $y=1/2$ to evaluate the
Wronskian of $u_{\scriptstyle {\rm {reg}}}$ and each one of those
functions, one has
\begin{eqnarray}
\lefteqn{\hspace{-1cm}\mathcal{W}[u_{\scriptstyle {\rm
{reg}}},u](y=1/2)=-\frac{1}{2^{\kappa/(2\alpha)+\mu}}
\Bigg(\left(\sum_{n=0}^{\infty}\frac{a_n}{2^n}\right)
\left(\sum_{m=0}^{\infty}\frac{(m+\mu)b_m}{2^{m-1}}\right)}
\nonumber \\
& & \hspace{5cm}\ +
\left(\sum_{n=0}^{\infty}\frac{(n+\kappa/(2\alpha))a_n}{2^{n-1}}\right)
\left(\sum_{m=0}^{\infty}\frac{b_m}{2^m}\right)\Bigg).
\label{cdos}
\end{eqnarray}
It can be checked numerically that the right hand side of
(\ref{cdos}) becomes zero if,
\[
{\rm{for}}\; n=0,1,2,\ldots, \quad
0<\frac{\kappa}{\alpha}=\left\{\begin{array}{ll} \lambda-1-2n
\quad {\mbox{\rm for even states,}} \\ \lambda-2-2n \quad
{\mbox{\rm for odd states.}}\end{array}\right.
\]

We arrive finally to the case of the Morse potential
\begin{equation}
V(r)=D(\exp(-2\alpha x)-2\exp(-\alpha x)), \qquad x=(r-r_0)/r_0,
\quad \alpha >0, \label{ctres}
\end{equation}
exactly solvable for angular momentum $l=0$. By introducing, as in
Ref. \cite{flu}, a new variable
\begin{equation}
y=\frac{2\gamma}{\alpha}\exp (-\alpha x), \label{ccuatro}
\end{equation}
and denoting
\begin{equation}
\beta^2=-\frac{2mEr_0^2}{\hbar^2},  \qquad
\gamma^2=\frac{2mDr_0^2}{\hbar^2},  \qquad \beta,\gamma>0,
\label{ccinco} \end{equation}
the Schr\"odinger equation becomes
\begin{equation}
y^2u^{\prime\prime}+yu^{\prime} + \left(-\frac{\beta^2}{\alpha^2}+
\frac{\gamma}{\alpha}y-\frac{1}{4}y^2\right)u = 0. \label{cseis}
\end{equation}
This equation presents a regular singular point at the origin and
an irregular one at infinity. The physical solution, however,
needs to be defined only between $y=0$, corresponding to
$x\to\infty$ ($r\to\infty$), and
$y=y_0\equiv(2\gamma/\alpha)\exp(\alpha )$, corresponding to
$x=-1$ ($r=0$). Such physical solution must be regular at $y=0$
and become zero at $y=y_0$. The solution regular at $y=0$ can be
given as a series
\begin{equation}
u_{\scriptstyle {\rm {reg}}}(y)=\sum_{n=0}^{\infty}a_n\,
y^{n+\beta/\alpha}, \qquad a_0\neq 0, \label{csiete}
\end{equation}
with coefficients obeying
\begin{equation}
n(n+2\beta/\alpha)\,a_n=-(\gamma/\alpha)\,a_{n-1}+(1/4)\,a_{n-2}.
\label{cocho}
\end{equation}
The other extreme of the interval of definition of the wave
function, $y=y_0$, is an ordinary point. There are two independent
solutions of the differential equation, both finite at $y=y_0$.
But only the linear combination of them becoming zero at that
point is physically acceptable. Let us call it $u^{(1)}$. Now,
following our procedure, we should impose the cancellation of the
Wronskian of $u_{\scriptstyle {\rm {reg}}}$ and $u^{(1)}$ at any
point of $[0,y_0]$. If we choose $y=y_0$, it becomes
\begin{equation}
\mathcal{W}[u_{\scriptstyle {\rm {reg}}},u^{(1)}](y=y_0) =
u_{\scriptstyle {\rm {reg}}}(y_0)\,u^{(1)\prime}(y_0),
\label{cnueve}
\end{equation}
and, since $u^{(1)\prime}$ cannot vanish at $y=y_0$, the
quantization condition reads
\begin{equation}
u_{\scriptstyle {\rm {reg}}}(y_0)=0, \label{cincuenta}
\end{equation}
an expression that could have been obtained trivially, without
having recourse to our method. It is not difficult to see that, if
one takes $a_0=1$ in (\ref{csiete}), one has
\begin{equation}
u_{\scriptstyle {\rm {reg}}}(y)=y^{\beta/\alpha}\exp(-y/2)\
_1\!F_1\left(\frac{1}{2}+\frac{beta}{\alpha}-\frac{\gamma}{\alpha},
1+2\frac{\beta}{\alpha};y\right), \label{quno}
\end{equation}
and the quantization condition (\ref{cincuenta}) coincides with
that given in Ref. \cite{flu}.

Unlike what happened in the case of potential (\ref{dos}), the
numerical convergence of the power series giving the solutions of
the Schr\"odinger equation in the three last examples is rapid
enough as to guarantee an accurate computation of the
eigenfunctions. For instance, in the case of the P\"oschl-Teller
potential, the series in (\ref{vocho}) can be used for $y\leq 1/2$
and that in (\ref{treinta}) for $y\geq 1/2$, the coefficients
$a_0$ and $b_0$ being determined by continuity at $y=1/2$ and
normalization in the interval $y\in [0,1]$.

 \ack

It is a great pleasure to dedicate this work to Prof. Rafael
Guardiola on occasion of his sixtieth birthday. Suggestions of two
anonymous referees have contributed to improve considerably a
first version of the paper. Financial support from Comisi\'{o}n
Inter\-mi\-nis\-te\-rial de Ciencia y Tecnolog\'{\i}a is
acknowledged.

\end{document}